\documentclass[journal abbreviation, manuscript]{copernicus}
\graphicspath{ {./figures/} }

\begin{document}
\nolinenumbers
\title{Extended Aerosol Optical Depth (AOD) time series analysis in an Alpine Valley: A Comparative Study from 2007 to 2023}


\Author[1][jochen.wagner@i-med.ac.at]{Jochen}{Wagner} 
\Author[1]{Alma Anna}{Ubele}
\Author[1]{Verena}{Schenzinger}
\Author[1, 2]{Axel}{Kreuter}

\affil[1] {Institute of Biomedical Physics, Medical University of Innsbruck,M\"{u}llerstra\ss e 44, 6020 Innsbruck, Austria}
\affil[2] {LuftBlick Earth Observation Technologies, Fritz-Konzert-Strasse 4, 6020 Innsbruck, Austria}




\runningtitle{Extended Aerosol Optical Depth (AOD) time series analysis in an Alpine Valley: A Comparative Study from 2007 to 2023}

\runningauthor{Jochen Wagner}



\firstpage{1}

\maketitle

\begin{abstract}
This study presents an extended analysis of aerosol optical depth at 501 nm (AOD) in the Alpine valley of Innsbruck, Austria, from 2007 to 2023, and offers a comparative analysis with the Alpine station of Davos, Switzerland. AOD is derived from ground-based sunphotometer measurements of direct spectral irradiance during daytime. The Davos Station is part of the AErosol Robotic NETwork (AERONET), a global network providing high quality, ground based remote sensing aerosol data and complies with the relevant requirements. The Innsbruck station does not belong to AERONET, but the AOD retrieval algorithm is very similar. Building upon previous research conducted until 2012, the presented study aims to provide a comprehensive understanding of the long-term trends and seasonal variations in aerosol characteristics in Central Alpine regions. We observed the typical mid lattitude annual cycle with a maximum in July and a minimum in December. The AOD trends per decade for both stations are declining,  -27.9  x $10^{-3}$  for Innsbruck and -9.9  x $10^{-3}$ for Davos.
\end{abstract}


\introduction  

The interplay between atmospheric aerosols and environmental dynamics has long been a subject of keen scientific interest, particularly in the context of climate change (\citet{Li2022}), air quality, cloud microphysics (\citet{Tiwari2023}) and ecological impacts (\citet{Zhou2021}). Aerosol Optical Depth (AOD) is a pivotal parameter in this domain, offering a quantifiable measure of aerosol concentration in the Earth's atmosphere. Satellite derived AOD with global coverage improves our knowledge on the distribution (\citet{Levy2009}). However, satellite retrievals face limitations due to their viewing geometry, where light traverses the atmosphere twice and reflects off the Earth's surface, complicating accurate measurement, whereas ground based remote sensing observations meet the World Meteorological
Organization (WMO) traceability requirements in more than 95\% of the measurements (\citet{Cuevas2019}) and allow robust trend analyses (\citet{Kazadzis2018}). High quality AOD time series are of special importance regarding climate observations (\citet{Kassianov2021}). This study aims to deepen our understanding of aerosol behavior in the Alpine valleys of Innsbruck, Austria, and Davos, Switzerland. Unfortunately, other stations from AERONET (\citet{Giles2019}) like Zugspitze and Bolzano have only very limited measurement series.

The Alpine region, characterized by its distinct topography and climatic conditions, presents a natural laboratory for studying aerosols (\citet{Ingold2001}). The complex interactions of local and regional meteorological patterns, coupled with anthropogenic influences, make this area particularly interesting for long-term environmental observations of aerosols (\citet{Lenoble2008}). In this context, the city of Innsbruck, a valley station in the centre of the Tyrolean Alps, and the high-altitude station of Davos in Switzerland, provide contrasting yet complementary settings for examining aerosol characteristics.

In Europe, strict environmental regulations and implementations of cleaner technologies since the late 20th century have significantly reduced aerosol emissions, while an upward trend has been observed in other regions (\citet{Yu2020}). This "global brightning" effect became evident since the 1980s. It seems, that this effect is still ongoing since many studies show a decreasing AOD in Europe over the last 20 years (\citet{Cherian2020}). Our research is anchored in the long history of aerosol studies in Alpine environments, notably extending the work of \citet{Wuttke2012} and drawing comparative insights from recent findings by \citet{Karanikolas2022}. By analyzing a 17-year AOD dataset, this study seeks to uncover the long-term trends and seasonal variabilities of aerosols in two Alpine valleys. The extended timeframe of our analysis, spanning from 2007 to 2023, allows for a detailed exploration of the temporal evolution of aerosol characteristics, contributing to a broader understanding of their role in regional and global climatic systems.

The significance of this study lies not only in its extended temporal scope but also in its contribution to the ongoing discourse on environmental and climatic changes. By examining the trends and patterns in AOD data, we aim to provide valuable insights into the underlying processes driving aerosol distribution and concentration in the Alpine region. This research holds valuable information for future environmental policies and strategies aimed at mitigating the impacts of atmospheric aerosols on climate, ecosystems, and human health.

\section{Methods}

\begin{figure*}[t]
\includegraphics[width=12cm]{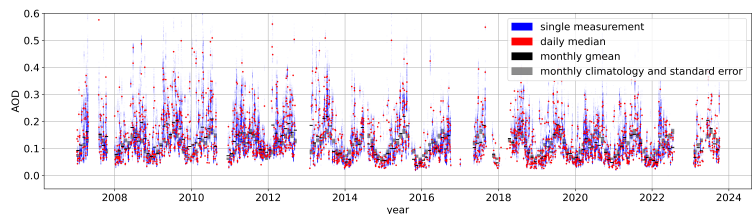}
\caption{AOD time series from Innsbruck. Individual measurements (minute intervals) are shown in blue, daily values in red and monthly averages (geometric mean) in black. In addition, the monthly climatologies standard errors are shown for comparison (see figure \ref{fig:figure4}).}
\label{fig:figure1}
\end{figure*}

\begin{figure*}[t]
\includegraphics[width=12cm]{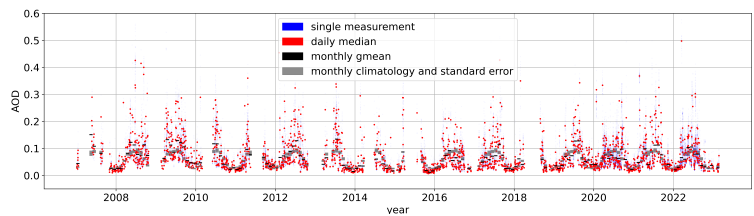}
\caption{AOD time series from Davos. Individual measurements (10 minute intervals) are shown in blue, daily values in red and monthly averages (geometric mean) in black. In addition, the monthly climatologies standard errors are shown for comparison (see figure \ref{fig:figure4}).}
\label{fig:figure1a}
\end{figure*}

Utilizing a robust dataset collected over 17 years in Innsbruck and Davos (see figure ~\ref{fig:figure1} and ~\ref{fig:figure1a}), we employ best practices (\citet{Sayer2019}, \citet{Weatherhead1998}) to analyse the AOD time series, focusing on identifying trends, patterns, and anomalies.
Both time series start in January 2007. The time series of Innsbruck ends in October 2023, whereas data from Davos were only available until February 2023. However, both time series can be compared very well because the measurements were generated with similar measuring devices \citet{Cuevas2019}. The temporal resolution in Innsbruck is 1 min and in Davos 10 min.  Furthermore the data availability with 48.6\% /42.1\% daily and 73.3\%/79.4\% monthly (see table \ref{table:first}), for Innsbruck and Davos respectively, is also very similar and remarkably high, given, that measurements are only possible when the sun is above the horizon and not obscured by clouds.
\begin{table*}[t]
\caption{The number of measurements of the datasets, the time period used and the number of days and months considered as valid with the percentage of valid days/months in brackets.}
\begin{tabular}{lccccccc} 
\hline  
Station & Lat & Lon & Elevation & Period & N & Valid Days & Valid Months \\ 
\hline 
Innsbruck & 47.26417 \degree \ N & 11.38569 \degree \ E & 620 m & 01/2007 - 10/2023 & 612962 & 2973/6117 (48.6\%) & 168/202 (83.2\%)   \\
Davos & 46.81281\degree \ N &  9.84369 \degree \ E & 1589 m & 01/2007 - 02/2023 & 78124 & 2479/5893 (42.1\%) & 154/194 (79.4\%) \\
\hline 
\end{tabular}
\label{table:first}
\end{table*}

 We calculated daily median values only for days with at least three measurements (also standard in AERONET processing). The daily AOD climatology was derived by calculating the median for each day of the year (see figure  ~\ref{fig:figure3} and ~\ref{fig:figure4}). From these values the monthly geometric mean was calculated if there were at least five valid days available. With this approach we calculated the monthly AOD from 168 out of 202 months (83.2\%) in Innsbruck and 154 out of 194 months (79.4\%) in Davos (table \ref{table:first}) . The study also deals with a comparative analysis, highlighting the similarities and differences in aerosol behavior between the two locations. One of the main aims of the work is to perform a trend analysis on the monthly time series. First, we deseasonalized the time series of the monthly AOD and applied linear fitting on the residuals. Additionally we calculated the trends for each month using ideally 17 values. Our findings reveal negative trends in AOD.

\begin{figure*}[t]
\includegraphics[width=12cm]{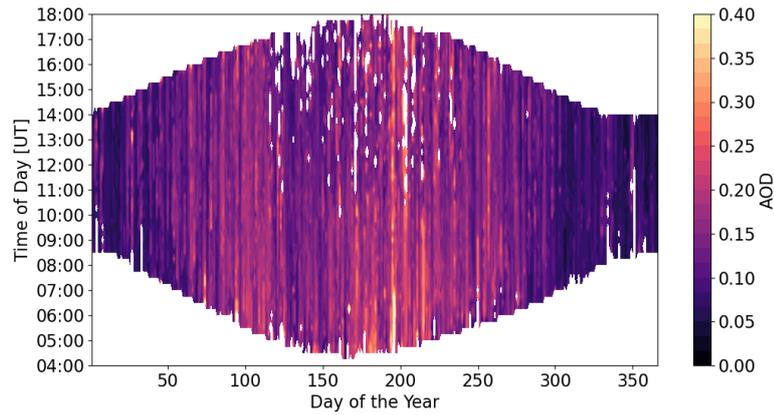}
\caption{Median AOD for each 15 min interval on each day of the year in Innsbruck. White areas indicate that there are no data available at these time points in the 17-year time series. }
\label{fig:figure2}
\end{figure*}

\begin{figure*}[t]
\includegraphics[width=12cm]{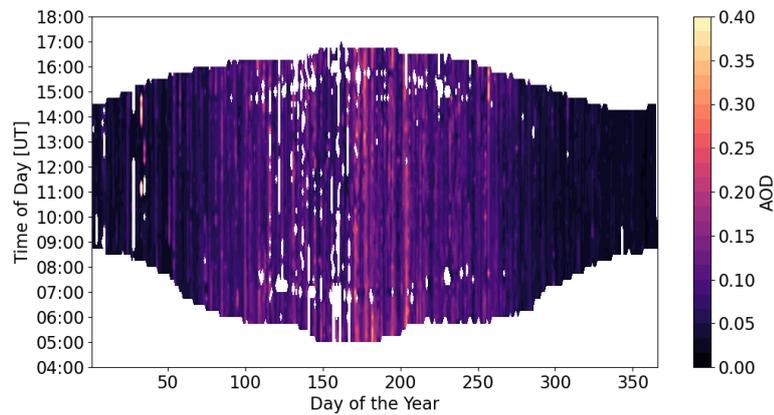}
\caption{Same like figure \ref{fig:figure2} for Davos}
\label{fig:figure2a}
\end{figure*}

\begin{figure}[t]
\includegraphics[width=8.3cm]{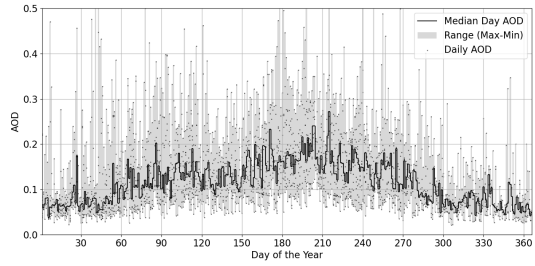}
\caption{Innsbruck daily 17 years AOD climatology. The median daily AOD is shown (black line) together with daily AOD (black dots) and the min-max range (grey background).}
\label{fig:figure3}
\end{figure}

\begin{figure}[t]
\includegraphics[width=8.3cm]{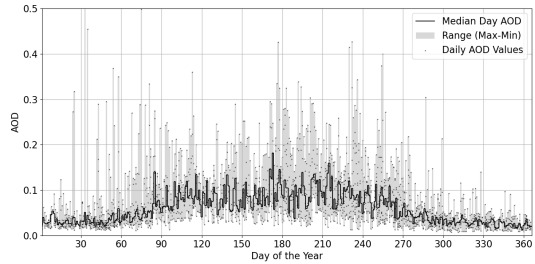}
\caption{Same like figure \ref{fig:figure3} for Davos}
\label{fig:figure3a}
\end{figure}

\subsection{Results}
A closer look at the two time series (figures ~\ref{fig:figure1} and ~\ref{fig:figure1a}) reveals the typical lognomal distribution of the AOD measurements (\citet{ONeill2000}). The highest value of 0.632 was measured in Innsbruck on 12 February 2010 and in Davos (0.864) on 2 February 2012 (both values outside the displayed y-range). The lowest value was observed on 2 November 2015 in Innsbruck (0.021) and on 7 November 2015 in Davos (0.007). Longer data gaps occur in both time series due to device failures or calibrations. Short gaps result from periods of bad weather. The typical annual variation is already recognizable, especially in the monthly averages. The average AOD in Davos (0.054) is about half as high as in Innsbruck (0.115).

The AOD is derived by measuring the direct irradiance of the sun. Therefore measurement errors and often correlate with the zenith and azimuth angle. Figures ~\ref{fig:figure2} and ~\ref{fig:figure2a} provide a good visual overview of the average annual and daily variation of the AOD at the two locations; no clear diurnal variation can be observed at either location. In Innsbruck, it is noticeable that there are many data gaps in the afternoon, especially in summer, which is probably due to convective clouds. In Davos, data gaps occur mainly in spring. This effect might occur due to more convective clouds in the afternoon during the melting period in Spring and early summer. In addition, there are particularly many data gaps here in the summer half-year with a solar zenith angle of approx. 15 degrees both in the morning and in the evening.

Due to the short time series (30 years is the standard for climatologies) and the data gaps due to cloudy days, the climatologies of the two stations on a daily basis (Figures ~\ref{fig:figure3} and ~\ref{fig:figure3a}) show (still) strong fluctuations. Nevertheless, the representation offers added value because the lognormal distribution becomes clear and extreme events can be quickly identified. This type of display is particularly suitable for daily data monitoring if the current data is also displayed in the graph.

The climatologies of the two stations on a monthly basis are a central result of this study. The annual mean value (geometric mean of the daily values) is 0.115 in Innsbruck and 0.054 in Davos. The month with the highest AOD is July (0.163/0.093) and the month with the lowest AOD is December (0.062/0.025) for Innsbruck and Davos respectively. The different altitudes and increased influence of human activities apparently only have an influence on the absolute value of the AOD, but not on the characteristic diurnal variation. The month of May is an exception. Here there is a local minimum in Innsbruck, while the month is unremarkable in Davos. This effect might be caused by differences in the annual cycle of the biosphere due to the difference in altitude between Davos and Innsbruck. However, further investigations are needed to prove this hypothesis.

We calculated the trend from the deseasonalized monthly AOD time series (figure ~\ref{fig:figure5}). For both stations a declining trend is obvious. For Innsbruck we calculated a trend of -27.9 x $10^{-3}$ and for Davos  -9.9  x $10^{-3}$. These trends are in line with the findings of \citet{Yang2020} and \citet{Wei2019}. Additionally we calculated the AOD trends per decade also for each month (table \ref{table:second}). The monthly trend calculations, due to the limited number of data points (11 - 16), are not yet very meaningful. Nevertheless, a fairly consistent pattern emerges again. With the exception of August and September in Davos all trends are negative. May shows the strongest negative trend in Davos and the third strongest negative trend in Innsbruck. August is the month with the least decrease in Innsbruck, or even a slight increase in Davos. In contrast, there are strong trend differences between Innsbruck and Davos in October and February.

\begin{figure*}[t]
\includegraphics[width=12cm]{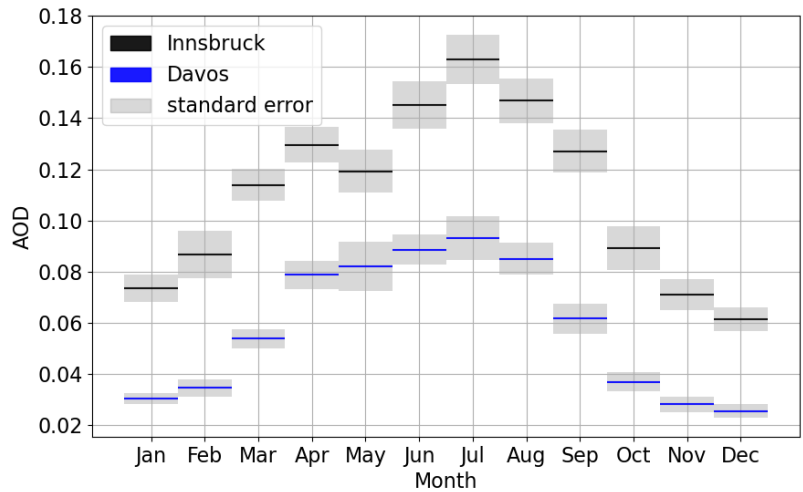}
\caption{Monthly AOD 17 years climatology with standard errors for Innsbruck and Davos}
\label{fig:figure4}
\end{figure*}

\begin{figure}[t]
\includegraphics[width=8.3cm]{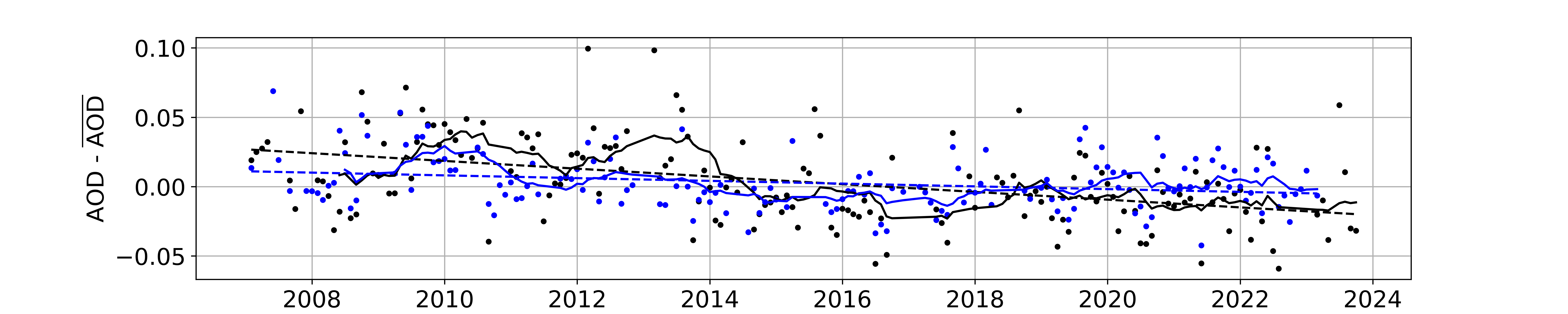}
\caption{Deseasonalized monthly AOD for Innsbruck (black dots) and Davos (blue dots). The 12 month running mean (Innsbruck  - black, Davos - blue) and the respective linear trends (dashed)}
\label{fig:figure5}
\end{figure}

\begin{table*}[t]
\caption{AOD trends per decade x $10^{-3}$ for each month in Innsbruck and Davos - number of valid months N in brackets}
\begin{tabular}{lcc} 
\hline  
Month & Innsbruck trend (N) & Davos trend (N)  \\ 
\hline 
1 & -25.9 (13) & -2.6 (12) \\
2 & -43.6 (15) & -1.9 (14) \\
3 & -18.9 (15) & -2.3 (14) \\
4 & -29.5 (16) & -18.1 (11) \\
5 & -38.6 (15) & -43.8 (11) \\
6 & -21.3 (16) & -18.7 (12) \\
7 & -31.4 (15) & -17.5 (12) \\
8 & -6.1 (16) & 6.4 (13) \\
9 & -29.0 (13) & -12.7 (15) \\
10 & -50.7 (11) & -10.5 (14) \\
11 & -20.2 (11) & 3.1 (13) \\
12 & -27.6 (12) & -0.6 (13) \\
all & -27.9 (168) & -9.9 (154) \\
\hline 
\end{tabular}
\label{table:second}
\end{table*}


\conclusions[Conclusions and Outlook] 
Overall, the results in AOD statistics for Innsbruck and Davos are remarkably consistent. The trends are as expected (\citet{Yang2020} and \citet{Wei2019}) and show, that the decline of AOD in the last 17 years can be observed in the lower and also the upper atmosphere. The observed decline is very likely due to a decline of anthropogenic emissions (\citet{Myhre2017}). It seems, that the local minimum in May in Innsbruck is becoming even more pronounced. Further investigations taking local emissions and land use changes into account are worthwhile. 

In summary, this study represents a significant step forward in our comprehension of aerosol climatology in the Alpine region, offering a nuanced understanding of the environmental statistics and long-term trends of aerosols in Innsbruck and Davos.



\dataavailability{The AOD measurements from Davos is available via Aeronet: https://aeronet.gsfc.nasa.gov/new\_web/photo\_db\_v3/Davos.html.  The AOD measurements for Innsbruck are available on request. } 













\authorcontribution{Jochen Wagner wrote the paper, performed most of the data analysis and made the figures. Alma Anna Ubele carried out part of the statistical analysis and updated the reference database. Verena Schenzinger has been responsible for the sunphotometer in Innsbruck and improved the AOD retrieval algorithm including cloud flagging. Axel Kreuter has operated the sunphotometer in Innsbruck for more than a decade. He was particularly involved in the planning of the paper and the figures.  } 

\competinginterests{The authors declare that they have no competing interests.} 


\begin{acknowledgements}
The authors acknowledge the Physikalisch-Meteorologisches Observatorium Davos / World Radiation Center (pmd/wrc) and AERONET-Europe/ACTRIS for long term operation and calibration and maintenance services of the CIMEL sunphotometer in Davos. 

This work was supported by the Medical University of Innsbruck.

\end{acknowledgements}

\bibliographystyle{copernicus}



\bibliography{references.bib}

\end{document}